\documentclass[prc,twocolumn,nofootinbib]{revtex4-2}

\usepackage{color}
\usepackage[dvipdfmx]{graphicx}
\usepackage{dcolumn}
\usepackage{bm}
\usepackage{braket}
\usepackage[dvipdfmx]{hyperref}

\usepackage[normalem]{ulem}  

\renewcommand\sout{\bgroup\color[rgb]{1,0.75,0.8} \ULdepth=-.5ex \ULset}

\renewcommand{\vec}[1]{\mbox{\boldmath $#1$}}

\begin{document}


\title{Role of triaxiality in deformed halo nuclei} 

\author{K. Uzawa, K. Hagino, and K. Yoshida}

\affiliation{%
 Department of Physics, Kyoto University, Kyoto 606-8502, Japan
}%

\date{Received\today}

\begin{abstract}
It is known that nuclear deformation plays an important role in inducing the 
halo structure in neutron-rich 
nuclei by mixing several angular momentum components. 
While previous theoretical studies on this problem in the literature 
assume axially symmetric deformation, we here consider non-axially symmetric 
deformations. 
With triaxial deformation, the $\Omega$ quantum number is admixed in a single-particle 
wave function, where $\Omega$ is the projection of the single-particle angular momentum 
on the symmetric axis, and the halo structure may arise even when it is absent with 
the axially symmetric deformation. 
In this way, the area of halo nuclei may be extended when triaxial deformation is considered.
We demonstrate this idea using a deformed Woods-Saxon potential for nuclei with neutron number N=13 and 43.
\end{abstract}


\maketitle


With the developments of radioactive isotope beam facilities worldwide,  
many experimental data on neutron-rich nuclei have been accumulated. 
Among them, the halo phenomenon, first observed in $^{11}\mathrm{Li}$ \cite{tanihata1985}, 
has attracted lots of attention \cite{zhukov1993,jensen2004,jonson2004,hagino2013,tanihata2013,nakamura2017,otsuka2020}. 
The halo nuclei are characterized by a spatially extended density distribution originated 
from weakly bound valence nucleon(s). Such structure can be probed, e.g., by  
interaction cross section measurements \cite{ozawa2001} and the Coulomb dissociation 
reactions \cite{nakamura2017}. 
It has been shown that 
the valence nucleons have to have 
a low orbital angular momentum, either $l=0$ or $l=1$, in order for the halo 
structure to emerge \cite{moller1992,sagawa1992,sagawa2015}.
With a higher angular momentum, the centrifugal barrier prevents 
the density distribution from extending, and the radius of a nucleus remains normal. 
As a consequence, $s$- and $p$- orbitals behave differently from 
orbitals with higher angular momentum, leading to a possible change in the shell 
structure \cite{hamamoto2001,hamamoto2010,hamamoto2012}.

Nuclear deformation significantly extends the region of halo nuclei \cite{Misu1996,hamamoto2004,zhou2010,hamamoto2017,hamamoto2021}. 
When a nucleus is deformed, the orbital angular momentum is not a good quantum 
number anymore, and they are admixed in a single-particle wave function. 
With an axially symmetric deformation, 
single-particle states with 
$\Omega^{\pi}=\frac{1}{2}^+,\frac{1}{2}^-$ and $\frac{3}{2}^-$ 
may then form a halo, where $\pi$ and $\Omega$ are parity and  
the projection of the total single-particle angular momentum $\vec{j}$ 
onto the symmetry axis, respectively,  
as these orbitals can contain the $s$-wave or the $p$-wave components. 
For instance, a $d_{3/2}$ orbital will couple to an $s_{1/2}$ orbital 
by nuclear deformation 
when $\Omega^\pi$ is equal to $\frac{1}{2}^+$, and a halo structure may arise even when the nucleus does not 
show it in the spherical limit. 
Moreover, it has been shown that the $s$-wave and $p$-wave components become dominant 
in a single-particle wave function when the separation energy is low \cite{Misu1996,hamamoto2004,yoshida2005}.
The experimental indications of the halo structure in 
$^{31}$Ne \cite{nakamura2009} 
and $^{37}$Mg \cite{takechi2014,kobayashi2014} 
are considered to be due to this mechanism, 
as the valence neutrons in these nuclei would occupy 
the $f_{7/2}$ orbital in the spherical limit \cite{hamamoto2010-2,minomo2011,minomo2011-2,sumi2012,urata2012,watanabe2014,urata2017,nakada2018,kasuya2021}. 

We notice that all the discussions on deformed halo nuclei 
have so far been based on axially 
symmetric deformation. In this paper, we extend this to triaxial deformation,
and clarify 
the role of triaxiality in deformed halo nuclei. 
When non-axially symmetric deformation occurs, $\Omega$ is no longer a good 
quantum number and several components of $\Omega$ are mixed in a wave function. 
For instance, even though an orbital with $\Omega^{\pi}=\frac{5}{2}^+$ does not form 
a halo structure with the axially symmetric deformation, the halo structure may 
arise in the presence of triaxial deformation, 
with mixing with an $\Omega^{\pi}=\frac{1}{2}^+$ component. 
This is in a sense similar to an extension of the halo region going from spherical 
symmetry to axial deformation. 

In order to demonstrate the role of triaxiality in deformed halo nuclei, 
let us consider a nucleus with neutron number 
$N$=43 . 
For this purpose, 
we employ a triaxially deformed Woods-Saxon (WS) potential, 
\begin{eqnarray}
V(\vec{r})=V_0f(\vec{r})-iV_{ls}r_0^2\nabla\cdot(\vec{s} \times \nabla)f(\vec{r}),
\label{WS}
\end{eqnarray}
with
\begin{equation}
f(\vec{r})=\frac{1}{1+\exp[(r-R(\theta,\phi))/a]},
\end{equation}
where the radius parameter $R(\theta,\phi)$ is given by 
\begin{equation}
R(\theta,\phi)=R_0\left[1+\sum_{\mu} a_{2\mu}Y^*_{2\mu}(\theta,\phi)\right]. 
\end{equation}
Here, $a_{2\mu}$ are the deformation parameters, $Y_{2\mu}$ are the spherical 
harmonics, and $\vec{s}$ is the spin operator. 
We use the parameters of 
$V_0=-42.86$ MeV, 
$r_0=1.27$ fm,
$R_0=5.20$ fm, $a$=0.67 fm, and $V_{ls}=-0.44V_0$.
For simplicity, we here consider only the quadrupole 
deformation. In the following, we express the deformation using the parameters $\beta$ and 
$\gamma$, instead of $a_{2\mu}$, which are defined as 
\begin{eqnarray}
a_{20}&=&\beta\cos\gamma, \\
a_{22}&=&a_{2-2}=\frac{1}{\sqrt{2}}\,\beta\sin\gamma, \\
a_{2\pm 1}&=&0.
\end{eqnarray}

We solve the Schr\"odinger equation for single-particle wave functions 
with the 3D mesh method \cite{bonche1985}. 
We impose the parity symmetry and the $z$-signature 
symmetry \cite{bonche1985, ogasawara2009} 
so that 
we need to consider only the 
positive coordinate region: 
$x_i =(i-1/2)\Delta r, y_j =(j-1/2)\Delta r,z_k =(k-1/2)\Delta r$ $(i,j,k=1,2,\cdots)$
with the mesh size of 
$\Delta r=0.9$ fm and 24 points for each direction.
We have confirmed that results are almost converged with this box size, unless 
the $s$-wave component dominates in the single-particle wave functions. 

Figure 1 shows the Nilsson diagram so obtained. 
In the spherical limit, the 43rd neutron occupies 
the 1$g_{9/2}$ orbit. With deformation, this level splits 
into five levels according to the value of $\Omega$. For a prolate deformation, the 
valence neutron 
occupies the orbital with $\Omega^\pi=\frac{3}{2}^+$, while 
the orbital with $\Omega^\pi=\frac{7}{2}^+$ in the oblate region. 
Notice that neither of these does form a halo structure. 

\begin{figure}
\includegraphics[width=8.6cm]{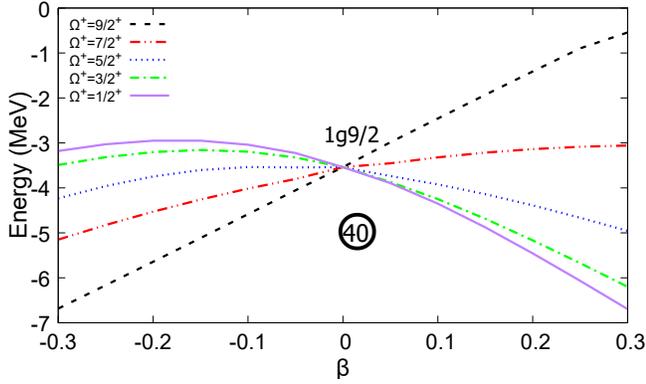}
\caption{Neutron levels as a function of the quadrupole 
deformation parameter $\beta$ obtained with an axially deformed 
Woods-Saxon potential. 
}
\end{figure}

Let us then take into account the triaxiality $\gamma$. 
Figs. 2 (a) and 2 (b) 
show the fraction of the s-wave component $P_s$ and 
the expectation value of the $z$-component of the total single-particle angular 
momentum, $\langle j_z\rangle$, respectively, 
for the single-particle level of the valence neutron. 
Different lines in the figures show the results of 
$\beta=0.3$ with several values of $\gamma$. 
To draw these figures, we 
vary $V_0$ in the Woods-Saxon potential to change the single-particle 
energy for the valence neutron. 
The fraction of the s-wave component is 
obtained by 
expanding the single-particle wave functions with 
the eigenfunctions of the spherical Woods-Saxon potential 
with $\beta=0$ in Eq. (\ref{WS}). 
We mention that $P_s$ does not reach 100\% in the region shown 
in the figure, partly because 
we impose a boundary condition that the single-particle wave functions 
vanish at the edge of the box. 
For $\gamma=0^\circ$ and $\gamma=60^\circ$, $P_s$ is zero since the 
$\Omega^\pi=3/2^+$ orbital has components with $j\geq 3/2$ and 
the $\Omega^\pi=7/2^+$ orbital has components with $j\geq 7/2$ so $s_{1/2}$ cannot 
be mixed. 
As the triaxial deformation 
parameter $\gamma$ increases 
from $\gamma=0^\circ$ or decreases from $\gamma=60^\circ$, 
the $s$-wave component gradually increases.  
Accordingly, 
$\langle j_z\rangle$ decreases and deviates from the half-integer values, 
as is shown in Fig. 2(b). 
These features are also seen in Fig. 3(a) and 3(b), which show $P_s$ and $\langle j_z\rangle$ respectively,
as a function of $\gamma$ at $S_n=0.2$ MeV.
One can see that 
with a triaxial deformation, 
the $s$-wave component increases significantly 
as the single-particle energy approaches 
zero, and a nucleus may form a halo structure.

\begin{figure}
\hspace*{-5.8cm}
\vspace*{-2cm}
\includegraphics[bb=126.0 36.0 486.0 706.0,scale=0.7]{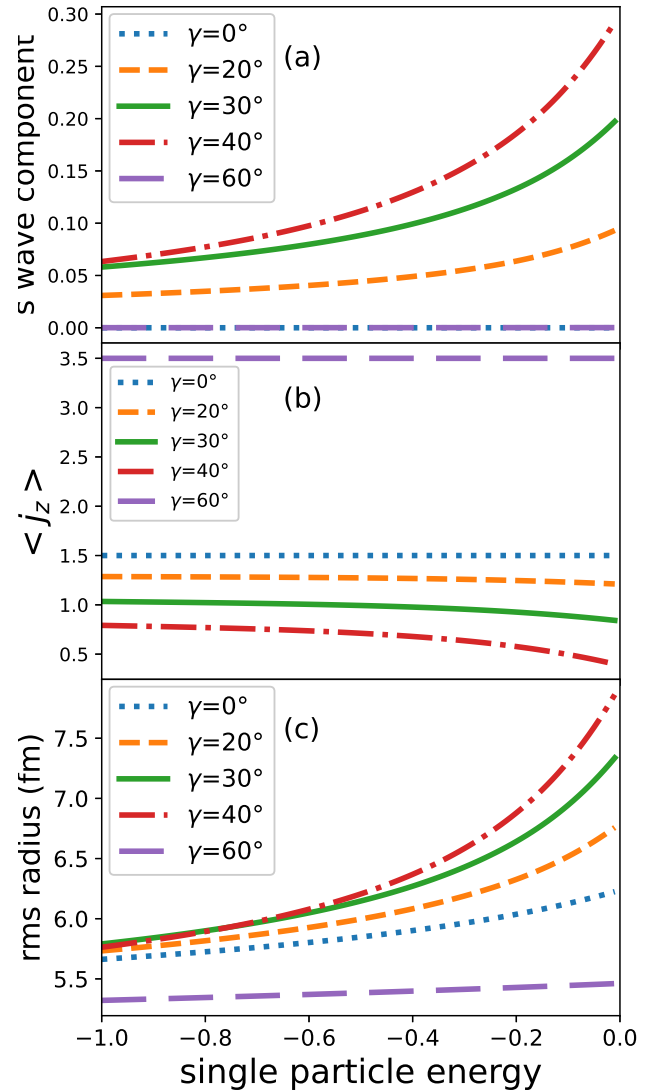}
\caption{The fraction of the $s$-wave component (Fig. 2(a)), the expectation 
value of the total single-particle angular momentum 
$j_z$ (Fig. 2(b)), and the root-mean-square (rms) radius (Fig. 2(c)) 
for the 43rd neutron orbital in a deformed Woods-Saxon potential. 
These are plotted as a function of the single-particle energy for several values of triaxiality
parameter $\gamma$ with $\beta=0.3$.
}
\end{figure}

\begin{figure}
\hspace*{-7.5cm}
\vspace*{-4cm}
\includegraphics[bb=134.925 149.46328125000005 527.075 742.53671875,scale=0.8]{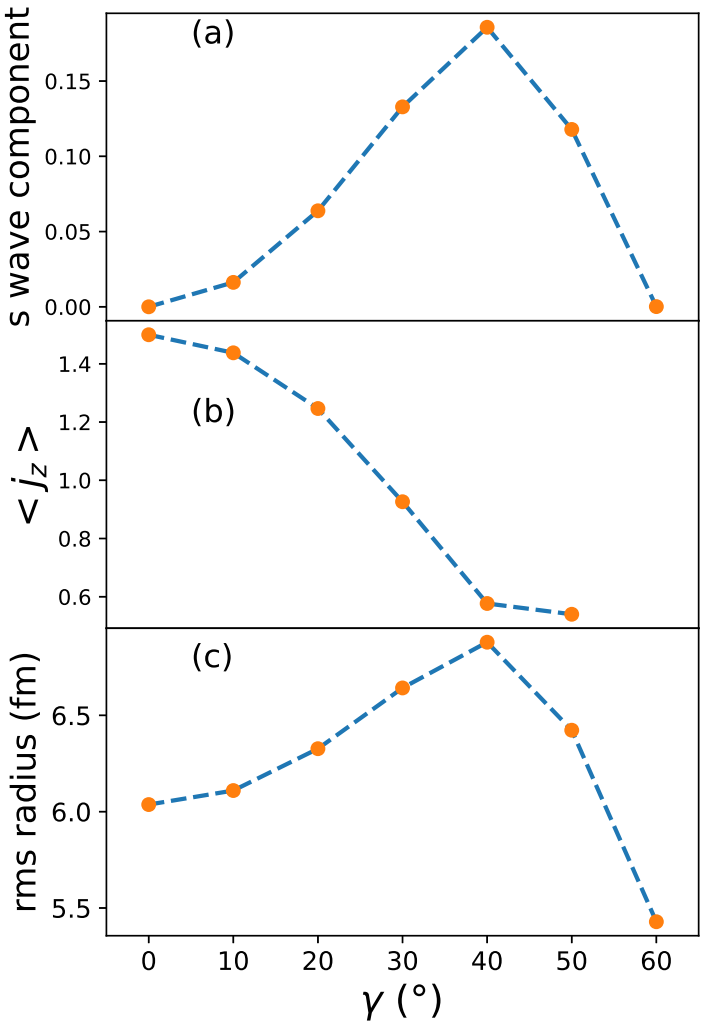}
  \caption{Similar to Fig. 2, but as a function of $\gamma$ for $S_n=0.2$ MeV.
  }
\end{figure}

The root-mean-square (rms) radius $\sqrt{<r^2>}$ of the valence neutron is plotted 
in Fig. 2 (c) as a function of the single-particle energy. 
As one expects from the behavior of the $s$-wave component, 
with the triaxial deformation the rms radius increases considerably as the single-particle energy approaches 
zero. 
Notice that this does not happen with axial deformation and triaxial deformation plays an essential role. 
In previous investigations of deformed halo nuclei, 
axial symmetry is assumed in most cases. Our work presented in this paper 
predicts that the region of halo nuclei can be extended if triaxial deformation emerges.

We next investigate briefly a nucleus with N=13, such as 
the $^{19}\mathrm{C}$ nucleus, which is considered to be a one-neutron 
halo nucleus  \cite{nakamura1999,ozawa2001,bazin1995,kanungo2016,wan2001, cao2018, 
sun2020,zhang2008,kanada2005}. 
Figure 4 shows the Nilsson diagram for this nucleus.  
To this end, we use the parameters of 
$V_0=-32.375$ MeV and $R_0=r_0A^{1/3}$ with $A=19$, while the 
other parameters are the same as the ones for Fig. 1. 
For the $^{19}$C nucleus, the valence neutron (i.e., the 13th neutron) occupies 
the orbital with $\Omega^{\pi}=\frac{1}{2}^+$ 
and $\Omega^{\pi}=\frac{3}{2}^+$ with a prolate and an oblate deformation, 
respectiely. 
This orbital can thus form a halo in 
the prolate region while it cannot have the halo structure in the oblate region. 
With triaxial deformation, both of these aspects are mixed together. 

\begin{figure}
\includegraphics[width=8.6cm]{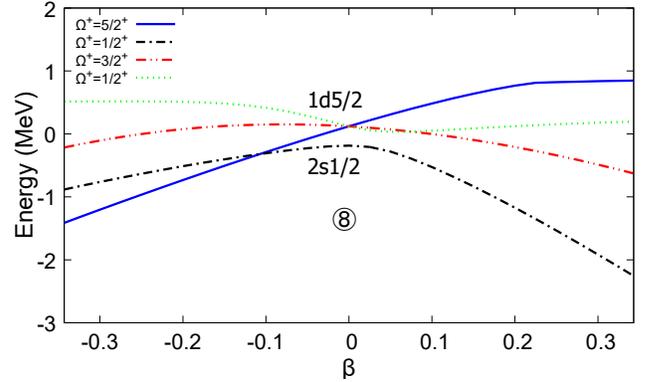}
\caption{A Nilsson diagram for neutron levels around N=13.
}
\end{figure}

Fig. 5 shows the rms radius as a function of the single-particle energy 
for $\beta=0.343$ with several values of $\gamma$. 
Here, $\beta=0.343$ is the optimized value for $^{18}$C obtained with the 
Skyrme-HF+BCS method \cite{zhang2008}. 
In the figure, the boxes with right diagonal lines and left diagonal lines 
denote the experimental 
radius $\sqrt{<r^2>}=5.5\pm0.3$ fm \cite{nakamura1999} and 
$\sqrt{<r^2>}=6.4\pm0.7$ fm \cite{kanungo2016}, respectively, together with 
the one-neutron separation energy $S_n=160\pm110$ keV obtained with the direct 
mass measurements \cite{audi1993}. 
The figure implies that 
the calculation with $\gamma=60^\circ$ underestimates the rms radius,
while the calculation with 
$\gamma=0^\circ$ clearly overestimates the rms radius.
By taking into account the triaxiality in the deformation, the halo component 
is modified in the wave function and 
the rms radius and the one-neutron separation energy are simultaneously 
reproduced with $\gamma=20^\circ-50^\circ$. 

\begin{figure}
\includegraphics[width=8.6cm]{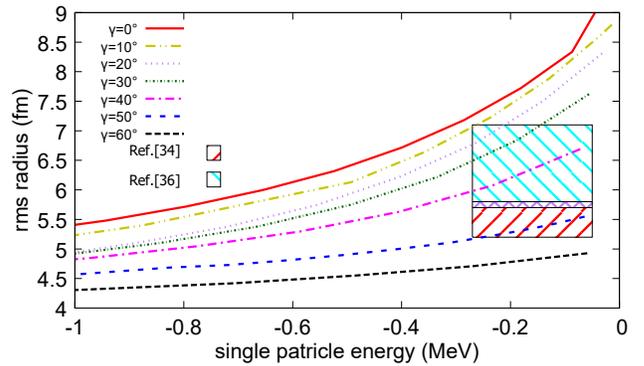}
  \caption{Relation between the single-particle energy of the valence neutron 
in $^{19}$C and the root-mean-square (rms) radius for several values of the triaxial 
deformation parameter $\gamma$. The deformation parameter $\beta$ is fixed to be 
0.343. The boxes with the right diagonal lines and the left diagonal 
lines show the experimental values of the rms radius 
from Refs. \cite{nakamura1999} and \cite{kanungo2016}, respectively, 
together with the empirical separation energy 
$S_n=160\pm110$ keV from Ref. \cite{audi1993}. 
}
\end{figure}

In summary, we have introduced a new perspective of deformed halo nuclei by considering triaxial deformation; we have pointed out that 
the mixing of configurations with different $\Omega$ quantum numbers 
with non-axially symmetric deformations 
may lead to the halo structure 
even when it does not appear 
with axially symmetric deformation. 
We have demonstrated this using neutron orbitals for the 43rd and the 13th neutrons 
in a deformed Woods-Saxon potential. 
The finding in this paper would enlarge the region of 
halo nuclei if triaxial deformation appears in neutron-rich 
medium-heavy and heavy nuclei. 
It would be an interesting future work to carry out a systematic study of deformed halo 
nuclei in a wide range of the nuclear chart by taking into account triaxial deformations.
In this connection, we mention that recent finite range droplet model (FRDM) calculations have indicated that medium-heavy nuclei near the drip lines deform triaxially in the wide region of the periodic table \cite{moller2008}. In addition, symmetry unrestricted HF+BCS calculations have shown that nuclei near the drip lines with N=60--80 and N=110--120 tend to be triaxially deformed \cite{ebata2017}.

\begin{acknowledgments}
We thank Nigel Orr for useful discussions on the direct mass measurements of $^{19}$C. 
This work was supported by JSPS KAKENHI
(Grant Nos. JP19K03824, JP19K03861, and JP19K03872).
The numerical calculations were performed
with the computer or computing facility 
at the Yukawa Institute for Theoretical
Physics, Kyoto University.
\end{acknowledgments}

\end{document}